\documentclass[aps,preprint,epsfig,multicol,nofootinbib,showpacs]{revtex4}
\usepackage{graphicx}
\usepackage{latexsym}

\begin{document}
\draft

\title{ $f^{-\gamma}$ Current Fluctuations in
Organic Semiconductors: \\ Evidence for Percolation
 }

\author{ A.~Carbone$^1$,  B.~K.~Kotowska$^{1,2}$, D.~Kotowski$^{1,2}$}

\affiliation{ $^1$ Physics Department, Politecnico
di Torino, C.so Duca degli Abruzzi 24, 10129 Torino, Italy \\
$^2$ Department of Physics of Electronic Phenomena, Gdansk
University of Technology, Narutowicza 11/12, 80-952 Gdansk,
Poland}



\begin{abstract}

The $f^{-\gamma}$ sloped current noise power spectra, observed
 in organic semiconductors,  have been interpreted within a  {\em
variable range hopping} mechanism of the fluctuations. The
relative current noise power spectral density ${\cal
S}(f)=S_I(f)/I^2$ exhibits a maximum
  at the {\em trap-filling transition }
between the {\em ohmic} and the {\em space-charge-limited-current}
regime [Phys.~Rev.~Lett., {\bf 95}, 236601, 2005]. Here, we
discuss the electronic conditions determining the crossover from
ohmic to space-charge-limited transport. These arguments shed
further light on the need to adopt a {\em percolative} fluctuation
picture to account for the competition between insulating and
conductive phases coexisting at the {\em transition}, where small
changes in the external bias lead to dramatic effects in the
fluctuations.

\end{abstract}


\maketitle  $f^{-\gamma}$ noise is an ubiquitous phenomenon
observed in many systems \cite{Kogan}. In spite of the intense
research effort, a general agreement on the origin of the
$f^{-\gamma}$ noise and the information, carried by it about the
underlying microscopic processes, has still to be achieved. In
homogeneous conductors, the relative spectral density of the
$f^{-\gamma}$ noise is independent of the voltage $V$ when the Ohm
law is obeyed. Furthermore, under the same assumptions,  an
increase (decrease) of the free charge carrier density $n$ results
in a monotonic decrease (increase) of the relative fluctuations.
The relative noise power spectral density of the current
$S_I(f)/I^2$, of the voltage $S_V(f)/V^2$, of the conductance
$S_G(f)/G^2$ and of the charge carrier density $S_n(f)/n^2$ are
indeed related by the following identities:

\begin{equation}
\label{dv} {\cal S}(f)=\frac{S_I(f)}{ I^2}=\frac{S_V(f)}{
V^2}=\frac{S_G(f)}{ G^2}=\frac{S_n(f)}{n^2 }\hspace{10pt}.
\end{equation}

Under the assumption of Poisson distributed fluctuations of the
free charge carrier density $n$, it is $\langle\delta n^2\rangle
\propto n$ and the relationship (\ref{dv}) can be written as:

\begin{equation}
\label{dn} {\cal S}\propto \frac{1}{n}f^{-\gamma}\hspace{10pt}.
\end{equation}

 This simple
proportionality fails to apply to inhomogeneous condensed matter
systems maintained away from thermal equilibrium by an external
excitation. The application of an electric, magnetic or photon
field, due the diverse conductive properties of the coexisting
phases and interfaces, causes the current paths to evolve
spatiotemporally with respect to quasiequilibrium conditions.
Similar phenomena occur in ferroelectrics, polymers and
copolymers, superconductors and magnetic semiconductors,
island-like metallic films, carbon-wax mixtures, polycrystalline
semiconductors only to mention a few examples \cite{Dagotto}. The
common feature shared by these systems is the transformation
undergone by the conduction patterns upon variation of a control
parameter in an otherwise quasi-homogeneous structure. The
modification of the conduction patterns upon an external bias
results in a current flow occurring through a mixed-phase
percolation process arising from a variable-strength competition
between multiphase metallic and insulating components. Due to the
strong localization and electron/hole interaction within the
disordered lattice, the transport in such systems is characterized
by phenomena as variable-range-hopping and polaronic effects.
 Both the amplitude and the
spectral characteristics of the noise are extremely sensitive to
the dynamics of the current paths upon the external excitation,
current fluctuations have been demonstrated to be helpful when
 the electronic properties of inhomogeneous systems have to be probed
\cite{Shklovskii,Carbone98,Carbone01,Carbone05,Pennetta,Bardhan,Chiteme,Planes,Li,Marley,Raquet}.
As a rule,  the behavior expected on the basis of the
Eq.(\ref{dv}) for homogeneous conductors, is not observed in
inhomogeneous systems. Relative noise power spectral densities
changing with the bias have been indeed reported. Furthermore, the
$1/n$ dependence of the relative noise lacks to occur when the
charge carrier density $n$ is changed. $1/n$
  deviations are for example obtained when polycrystalline
photosensitive materials are irradiated.  The photon flux-
preferentially and disorderly - increases the conductivity of the
regions where photosensitive defects are located,  resulting in
the formation of coexisting paths having different conductivities
\cite{Carbone98,Carbone01}. Another feature often observed in
inhomogeneous systems is the non-gaussianity of the noise traces.
All the mentioned issues are diverse aspects of an unique problem:
the charge carrier transport takes places across narrow conductive
paths, with a volume which is only a small part of the whole
conductor and with always less "fluctuators" involved in the
stochastic process.
\par $f^{-\gamma}$ fluctuations have been recently observed in
thin films of pentacene  and tetracene.  Pentacene and tetracene
are small-weight organic molecules formed respectively by five,
$C_{22}H_{14}$, and four, $C_{18}H_{12}$, benzene-like rings
\cite{Carbone05}. The study of noise in organic insulators is
interesting  at least for two reasons: (i) a complete
understanding of the
 mechanisms underlying the charge carrier transport in  organic small chain and polymers  has
 not yet
been fulfilled and still many unsolved issues remain; (ii)  the
deployment of organic and polymeric materials in the electronic
industry requires a detailed insight into the dynamics other than
into the time-averaged properties of the charge carrier transport
\cite{Lampert,Parris,Fishchuk,Deboer,Yang,Knipp,Lang,Muller,Kang,Mihailetchi,Lebowitz}.
\par
We have reported that the relative
 power spectral density ${\cal S}{(f)}$ observed in
 polycrystalline polyacenes is consistent  with steady fluctuations  of thermally generated and of injected
charge carriers, respectively in {\em ohmic} and in {\em
space-charge-limited-current} (SCLC) regime. The relative noise
suddenly increases at the {\em trap filling} region at
intermediate voltage.  The peak has been estimated within a simple
 percolation model of the fluctuations as a consequence of the imbalance between empty and
filled traps.
 The {\em ohmic}
transport is governed by thermally activated charge carriers with
the deep traps almost completely empty. In SCLC regime, the
transport is governed by the injected charge carriers, controlled
by space-charge, with the deep traps almost completely filled. The
intermediate voltage region, the {\em trap-filling transition}
(TFT), is characterized by the coexistence of a conductive  and an
insulating phase, corresponding respectively to the empty and
filled traps. The system can be viewed as a two-components
continuum percolative medium. The material, initially in the
quasi-homogeneous ohmic phase, becomes populated by insulating
sites as the voltage increases. The conductivity patches become
extremely intricate owing to the inhomogeneous distribution of
trapping centers, whose occupancy randomly evolves as the Fermi
level moves through the trap level. The system is in a strongly
disordered state, due to the nucleation of insulating patterns
inside the conductive medium. The relative noise intensity ${\cal
S}(f)$ of the system undergoing the trap-filling-transition
exceeds
 that of the same system
when one of the two phases prevails. The increase of fluctuations
has its origin in the greatly disordered distribution of local
fields compared to the almost uniform distribution characterizing
 the ohmic and the SCLC regimes. The resistance $R$ and the excess
noise ${\cal S}(f)$  observed in a percolative system are
described by means of the relationships
\cite{Stanley,Rammal,Tremblay,Balberg}:

\begin{equation}\label{R}
    R=\frac{1}{I^2}\sum_{\alpha} r_{\alpha} i_{\alpha}^2
\end{equation}

\begin{equation}
\label{percolativenoise} {\cal
S}(f)={s}_{\Omega}(f)\frac{\sum_\alpha i^4_\alpha}{(\sum_\alpha
i^2_\alpha)^2} \hspace{10pt},
\end{equation}
where $i_\alpha$ is the current flowing through the elementary
resistances $r_{\alpha}$ forming the network. I is the total
current, ${s}_{\Omega}(f)$ indicates the noise spectral density of
the conductive elements of the network. The resistance $R$ and the
excess noise ${\cal S}(f)$ progressively increase as the
conductive matrix becomes sparse according to the relationships:
\begin{equation}\label{Rc}
    R\propto (\Delta \phi)^{-t}
\end{equation}
\begin{equation}
\label{percolativeS}
 {\cal S} \propto (\Delta \phi)^{-k}
\end{equation}
where $\phi$ represents the conductive volume fraction, while $t$
and $k$ are critical exponents  depending on the structure,
composition and conduction mechanism of the percolative system.
The $t$ and $k$ values depend on the model (e.g. lattice, random
void (RV), inverted random void (IRV)) adopted to describe the
system \cite{Kogan,Balberg}.
\par An
expression of ${\cal S}(f)$ in terms of physical observable
depending upon the external drive has been worked out in
\cite{Carbone05}. During trap-filling, the conductive site
fraction of the network is reduced proportionally to the
difference between the free and trapped charge carrier densities.
It is:

\begin{equation}
\label{deltafi}
 \Delta \phi \propto \frac{n-n_t}{N_v}
\end{equation}
 where $n$ and $n_t$
are respectively the free and trapped charge carrier densities,
${N_v}$ is the total density of states, coinciding with the
molecular density for narrow band materials as polyacenes. It is
convenient to write the Eq. (\ref{deltafi}) as:

\begin{equation}
\label{deltafi2}
 \Delta \phi \propto \frac{n}{N_v}\left({1-\frac{n_t}{n}}\right)
\end{equation}
By comparing Eq.~(\ref{deltafi}) with Eq.~(\ref{dv}), it follows
that the noise, exceeding the level that would be expected for an
homogeneous conductor with comparable density of free charge,
arises from the term $(1-{n_t}/n)$, i.e. from the imbalance
between free and trapped carriers, causing the departure from the
quasiequilibrium ohmic conditions. Assuming for simplicity a
discrete trap level, it is $n=N_v \exp[-(E_v-E_F)/kT]$ and
$n_t={N_t}/{\{1+g^{-1}\exp[-(E_F-E_t)/kT]\}} \simeq {2
N_t}\exp[(E_F-E_t)/kT]$, $E_F$ being the quasi-Fermi level, $g$
the degeneracy factor of the trap, $N_t$ is the total density of
deep traps and the other quantities have the usual meaning
\cite{Lampert}. The resistance $R$ and the excess noise ${\cal S}$
diverge at the percolation threshold $\phi_c$ according to the
relationships:
   $ R\propto (\phi -\phi_c)^{-t}$
and $ {\cal S} \propto (\phi -\phi_c)^{-k}$. The percolation
threshold $\phi_c$ and the onset of breakdown  were obtained as a
consequence of additional traps progressively formed by  bias or
thermal stress, that through the increase of the trap density
$N_t$, enhance the unbalance between free and trapped charge
carriers $n_{t}/n$ \cite{Carbone05}.

\begin{figure}

\includegraphics[width=8cm,height=6cm,angle=0]{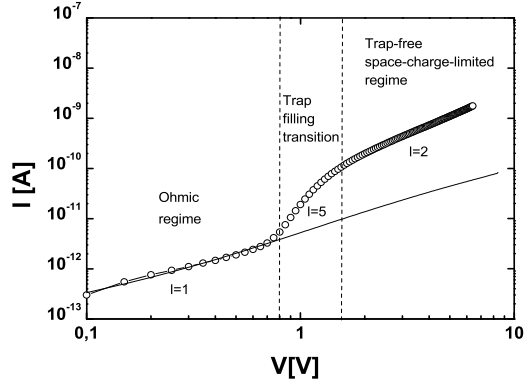}
\caption{\label{IV} Log-log plot of the I-V characteristics for a
tetracene sample Au/Tc/Al with $L=0.65 \mu m$ (circles) exhibiting
the typical behavior of steady-state SCLC in materials with deep
traps (ohmic regime$\Rightarrow$trap-filling
transition$\Rightarrow$SCLC regime).  The solid line corresponds
to the fully ohmic behavior over the entire voltage range }
\end{figure}

\begin{figure}
\includegraphics[width=8cm,height=6cm,angle=0]{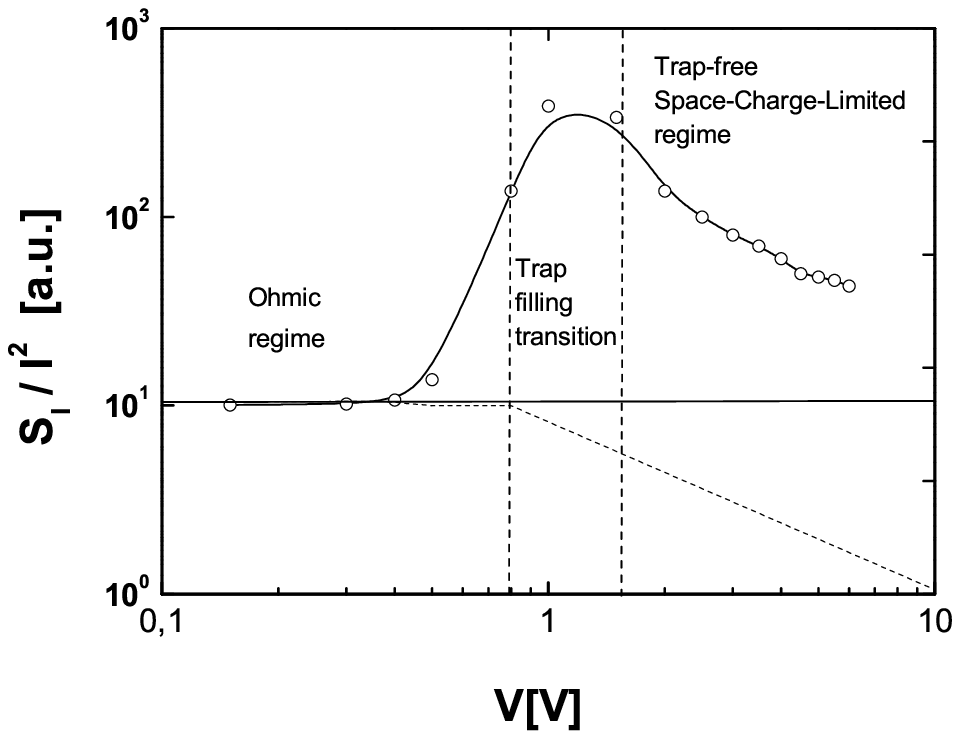}
\caption{\label{Noiseall} Log-log plot of the voltage dependence
of the relative fluctuation power spectral density at frequency
$f=20Hz$ for the Au/Tc/Al sample (circles). The horizontal solid
line, given for reference purpose,  corresponds to the behavior
expected for ohmic transport over all the voltage range. The dot
line, given for reference purpose, represents the decreasing
behavior that would be observed in an homogeneous system
undergoing a crossover from ohmic to SCLC conduction.}
\end{figure}

 \par The organic material has been figured as a binary phase system where the
transport is confined to narrower and narrower conductive paths
worn away by larger and larger insulating regions as the voltage
increases. In the remainder of the paper we will add further
insights into the percolative fluctuation model by arguing on the
relationships
  between thermal and injected charge carrier densities. These
  relationships will be then used to show that the observed
  behavior could not have been deduced as linear superposition of the
  fluctuations of the two diverse components, therefore evidencing of the percolative mechanism of the
  fluctuations.  In summary, this study demonstrates that an investigation accounting for the
  statistical properties, namely the charge carrier fluctuations, might be a key to unravel many puzzling
  phenomena related to the physics of organic and polymeric materials.
 Pure organic materials could be considered {\em perfect
insulators}, i.e. materials that, under an applied voltage, carry
a negligible current associated to the injection of charges
according to a mechanism, similar to the emission from a
thermionic cathode into vacuum, known as SCLC, analytically
described by the {\em Child law}. Compared to vacuum, the
description of the space-charge-limited current in organic and
inorganic solids requires to keep into account the complications
arising from the electron-phonon interactions. Additional issues
arise if chemical impurities and structural imperfections are to
be taken into consideration as it always happens in real as
opposed to ideal organic insulators. As a matter of fact, the
onset of charge injection under SCLC conditions from the
  electrode critically depends on
 the presence of deep and shallow energy states related to the unavoidable defects at the metal-organic interface
 and in the
 bulk. The curve in figure~\ref{IV} exemplifies the typical
behavior of the steady-state SCLC in materials with deep traps.
The low-voltage region, with slope $l\approx 1$, corresponds to
the {\em ohmic} regime with a current density described by:
\begin{equation}\label{ohmic}
    J_{\Omega}= \frac{e \mu n V}{L}  \hspace {5pt}.
\end{equation}
The high-voltage region, with slope $l \approx 2$, corresponds to
the trap-free space-charge-limited-current regime, obeying the
{\em Mott-Gurney law}:

\begin{equation}
\label{SCL}
 J_{\rm SCLC}=\frac{9 \epsilon\epsilon_0\mu \Theta {V^2}}{8 L^3}
\end{equation}
 where $\Theta$ is the trapping parameter and the other quantities have
the usual meaning. The current-voltage characteristic in the
intermediate voltage region, the trap-filling transition (TFT), is
analytically described by the {\em Mark-Helfrich law}
\cite{Lampert}:

\begin{equation}
\label{TFT}
 J_{\rm TFT}=N_v\mu e^{1-l}\left[\frac{\epsilon l}{N_t(l+1)}\right]^l \left(\frac{2l+1}{l+1}\right)^{(l+1)}\frac{V^{l+1}}{L^{2l+1}}
\end{equation}

The current crossing the material under SCLC regime is carried by
the injected carriers $n_{inj}$, depending on $V$ according to:

\begin{equation}
\label{ninj} n_{inj}=\frac{\epsilon \epsilon_0}{e L} \ V
\end{equation}

The ohmic regime described by the Eq.~(\ref{ohmic}) dominates up
to a voltage, the threshold voltage $V_t$, where the injected free
charge carrier density becomes comparable to the thermal
concentration. Furthermore, one can assume that the injected
charges have completely filled the traps at  $V_t$, as expressed
by the relationship:
\begin{equation}
\label{nt} n_{inj}(V_t)\approx n_{t}(V_t)
\end{equation}
  At the  crossover from ohmic to SCLC regime, the charge carrier transit
  time $\tau_t=L^2/\mu_e V$
   and  the dielectric relaxation time $\tau_r={\epsilon \epsilon_0}/e n
   \mu$ become comparable, i.e. $\tau_t=\tau_r$. This further  condition provides a relationship for the threshold voltage
    $V_t = {N_t e L^2}/{\epsilon \epsilon_0}$.
    \par
The sudden increase exhibited by the current at the trap-filling
transition is customarily estimated by assuming that it takes
place over a voltage range $\delta V_{t}=V_{tf}-V_{t}$ which is
proportional to the threshold voltage $V_t$, i.e $\delta V_{t}=c
V_t$ and $V_{tf}=(c+1)V_t$ (for the curve shown in Fig. (\ref{IV})
it is $V_t\simeq 0.8V$, $V_{tf}\simeq 1.6V$, $c\simeq 1$).
Therefore, because of the proportionality between the injected
charges and the voltage [Eq.~(\ref{ninj})], the change of injected
charges over $\delta V_{t}$ is given by:

\begin{equation}
\label{ninj2} \delta
n_{inj}=n_{inj}(V_{tf})-n_{inj}(V_{t})=(c-1)n_{inj}(V_{t})
\hspace{10pt}.
\end{equation}
\noindent Since, on the average, the traps are completely filled
at $V_t$, the additional charge $n_{inj}$ must all appear in the
valence band (respectively in the conduction band for electron
conduction). The trap filling process is thus accompanied by a
conductivity change $\delta \sigma = e \mu \delta n_{inj}$, given
by:

\begin{equation}
\label{deltaG} \delta \sigma=\frac{\partial \sigma}{\partial
    V_t} \delta V_t = e \mu (c-1) n_{inj}(V_t) \hspace{10pt},
\end{equation}
\noindent
 that accounts for the current increase. The latter
relationship could be obtained also by using the {\em regional
approximation} approach \cite{Lampert}, that is however beyond the
scope of the present discussion. \par The relationships
(\ref{ninj}-\ref{deltaG})  will be now used to argue  on the need
of the percolation picture for the current fluctuations observed
at the trap-filling transition. If the transport process would
occur in an homogeneous system, with unchanged volumes of the
conductive and insulating phase, the relative noise ${\cal S}(f)$
could have been evaluated using the Eq.(\ref{dv}). The increase of
$n$, due to the contribution of the injected $n_{inj}$ charge
carriers, would thus result in a decrease of the relative noise.
The hypothetical behavior that would be obtained for the
homogeneous conductor is represented by the dot line in
Fig.(\ref{Noiseall}). This behavior is of course opposed to the
noise increase experimentally observed.
 The charge carrier trapping results in the formation of insulating
regions and forces the current to flow across narrower and
narrower conductive paths. The dynamic evolution of such a complex
structure, where the charge carrier transport takes place, is
clearly deduced from the occurrence of a relative noise peak.  It
could not have been {\em a priori} deduced  from the $I-V$
characteristics, that, as we have argued above, will lead to the
opposite result. The high sensitivity of the noise to the dynamic
of the underlying conductive patterns is well-known: the current
noise is proportional to the fourth moment while the conductance
is proportional only to the second moment of the current
distribution. The behavior exhibited by the organic system is
analogous to other complex systems that spontaneously tend to form
patterns
 with varying size and scale over time (self-organization). In
such systems, the emergence of an aspect (e.g. giant
magnetoresistance, high-$T_c$ superconductivity, metal-insulator
transition) could not be derived as a linear superposition of the
properties of the single constituents. \par Let us finally discuss
the extent validity  of the percolation fluctuation model
\cite{Kogan,Rammal,Tremblay} with particular regards to the
organic materials. As already stated, the relationships
(\ref{R}-\ref{percolativeS}) apply only to  ohmic and trap-filling
regions of the I-V characteristics, i.e. in the subnetwork where
transport mostly takes place by continuum metallic conduction. As
the trap-free SCLC regime is approached, a condition guaranteed by
the onset of the 2-sloped $I-V$ curve, the noise mostly arises
from a mechanism different than the
 charge carriers density and/or mobility fluctuations.
The current is governed by the process of  injection and thus the
noise is very likely originated by fluctuations of the probability
of  emission or tunnelling across a potential barrier. The
mechanism  is, for certain aspects, analogous to the shot noise,
$S_{\rm shot}=2eI\Gamma $, observed in vacuum tubes or solid state
junctions, where the space charge, built-up in the interelectrode
region, acts with a negative feedback effect on the fluctuations
\cite{Carbone05}. Such a noise mechanism could indeed explain the
decrease of the relative fluctuations observed when the trap-free
space-charge-limited conduction is fully achieved. A more
complicated
 relationship than $S_{\rm shot}=2eI\Gamma $ should be however expected, due
 to the correlated hopping events and polaronic
effects in organic semiconductors. Another limit is represented by
the fact that the
 percolation process through a two-component disordered system, made of
ohmic and SCLC elements, is not exactly described by the
Eqs.(\ref{R}-\ref{percolativeS}).  In fact, these
 expressions contemplate the case of a
metal-insulator mixture with the insulator carrying no current at
all \cite{Rammal}. The two-components percolative theory
 is not adequate to the description of the present
case, as well.  In fact, a matrix formed by two components having
different ohmic conductivities,  with the current fluctuations
arising from the same mechanism  is dealt with in \cite{Tremblay}.
The analytical treatment of the percolative fluctuations in a
disordered space-charge-limited conductor would require that two
different mechanisms of noise related respectively to the ohmic
and to the SCLC phases,  are taken into account.
    \par
  To conclude, the investigation of current noise in
  polycrystalline polyacenes has revealed: (i) a {\em steady-state ohmic fluctuation
regime} at low-voltage; (ii) a {\em critical fluctuation regime},
at the trap filling transition, interpreted within a continuum
percolation model; (iii) a {\em steady-state
space-charge-limited-current fluctuation} regime at high voltage.
The current noise depends on the topological disorder due to
inhomogeneously distributed traps in the layer and thus it can
provide clues  in a microscopic description of transport in
organic materials.  These systems are characterized by distributed
thresholds for conduction, due to local inhomogeneous traps and
barriers of varying strength. The dynamical processes by which
such systems undergo a transition from order to disorder driven by
an external bias, are issues of general interest
 well beyond the field of organic electronics.
 The cooperative effect of charge, spin, lattice or
 impurity interactions, that could not be predicted on the basis of the {\em average}
superposition of the elementary constituents, is at the origin of
the giant fluctuations in such systems.  The relevant implication
is that the observed behavior  has to be considered as an {\em
emergent phenomenon}, within the broader context of the electronic
complexity.
 \par
Future issues to be investigated include experiments to observe
{\em space-charge-limited
 current
 fluctuations} in diverse materials and operative conditions.
 The development of a theoretical model of the percolative fluctuations, including two different noise components, will also represent
  an advancement in this field.
\par We acknowledge  financial support by the
 Italian Ministry for Foreign Affairs (MAE), under the contract
 22-FI-2004-2006.

\end{document}